\documentclass[journal]{IEEEtran}
\usepackage{amsmath}
\usepackage{graphicx}
\usepackage{amsfonts}
\usepackage{amsmath,epsfig}
\usepackage{stfloats}
\usepackage{amssymb}
\usepackage{cite}
\usepackage{algorithm}
\usepackage{algorithmic}

\begin{document}
%
\title{Joint Source and Relay Design for  Multi-user MIMO  Non-regenerative Relay Networks with Direct Links }

\author{ Haibin~Wan, and Wen~Chen,~\IEEEmembership{Senior~Member,~IEEE}
\thanks{Copyright (c) 2012 IEEE. Personal use of this material is permitted. However, permission
to use this material for any other purposes must be obtained from
the IEEE by sending a request to  pubs-permissions@ieee.org.}
\thanks{Manuscript received July 20, 2011; revised October 22, 2011 and March 1, 2012; accepted April 15, 2012. The associate editor coordinating the review of this paper
and approving it for publication was Hsiao-Feng Lu.}
\thanks{H.~Wan and W.~Chen  are with Department of Electronic Engineering, Shanghai Jiao Tong University,
China; H.~Wan is also with School of Physics Science and Technology,
Guangxi University,
China; W.~Chen is also with the SKL for ISN, Xidian University~(e-mail:\{dahai\_good;wenchen\}@sjtu.edu.cn)}

\thanks{This work is supported by national 973 project~\#2012CB316106,
by NSF China \#60972031 and \#61161130529, by national 973
project~\#2009CB824904, by national key laboratory project
\#ISN11-01, and by Foundation of GuangXi University~\#XGL090033.}
 }
\markboth{IEEE Transactions on Vehicular Technology ,~Vol.~X, No.~XX,~2012}%
{Shell \MakeLowercase{\textit{et al.}}: Bare Demo of IEEEtran.cls for Journals}

\maketitle

\begin{abstract}
In this paper, we investigate  joint source precoding matrices and
relay processing matrix  design for  multi-user multiple-input
multiple-output~(MU-MIMO) non-regenerative  relay  networks in the
presence of the direct source-destination~(S-D) links. We consider
both  capacity and mean-squared error~(MSE) criterions subject to
the distributed power constraints, which  are nonconvex and
apparently have no simple solutions. Therefore, we propose an
optimal source precoding matrix structure  based on the
point-to-point MIMO channel technique, and a new relay processing
matrix structure under the modified  power constraint at relay node,
based on which, a nested iterative algorithm of jointly optimizing
sources  precoding and relay processing  is established. We show
that the capacity based optimal source precoding matrices share the
same structure with the MSE based ones. So does the optimal relay
processing matrix. Simulation results demonstrate that the proposed
algorithm outperforms  the existing results.
\end{abstract}
\IEEEpeerreviewmaketitle

\section{Introduction}
\IEEEPARstart{R}{ecently}, MIMO relay network has attracted
considerable interest from both academic and industrial communities.
It has been verified that wireless relay can increase  coverage and
capacity of the wireless networks~\cite{2004-Relay-Based}.
Meanwhile, MIMO techniques can provide significant improvement for
the spectral efficiency and link reliability in scattered
environments because of its multiplexing and diversity
gains~\cite{2002-DMT}. A MIMO relay network, combining the relaying
and MIMO techniques, can make use of both advantages to increase the
data rate in the network edge and extend the network coverage. It is
a promising technique  for the next generation's wireless
communications.

The capacity of MIMO relay network has been extensively investigated
in the
literature~\cite{2005-BoWang,2011-WZJ01,2005-Caleb,2006-Helmut,2012-HBwan-WCL}.
Recent works on MIMO non-regenerative relay  are focusing on how to
design the source precoding matrix  and relay processing matrix.
For a single-user MIMO relay network, an optimal  relay processing
matrix which maximizes the end-to-end mutual information is designed
in~\cite{2007-Xiaojun} and~\cite{2007-Olga} independently, and the
optimal structures of jointly designed source precoding matrix and
relay processing matrix are derived in~\cite{2006-ZhengFang}.
In~\cite{2008-Guan} and~\cite{2008-Behbahani}, the relay processing
matrix to minimize the mean-squared error~(MSE) at the destination
is developed. A unified framework  to jointly optimize the source
precoding matrix and the relay processing matrix is established
in~\cite{2009-Rong-A-Unified}.  For  a multi-user single-antenna
relay network, the optimal relay processing is designed to maximize
the system
capacity~\cite{2007-Weng,2008-Chae-fixedRelay,2010-Gomadam}.
In~\cite{2010-YuanYu}, the optimal source precoding matrices and
relay processing matrix are developed in the downlink and uplink
scenarios of an MU-MIMO relay network without considering S-D links.
There are only a few works considering the direct S-D links.
In~\cite{2009-Rong-Direct} and~\cite{2010-Rong-SDirect}, the optimal
relay processing matrix is designed based on MSE criterion  with and
without the optimal source precoding matrix in the presence of
direct links, respectively.
However, for a relay network with direct S-D links, jointly
optimizing the source precoding matrix and the relay processing
matrix based on capacity or MSE is much difficult, especially for an
MU-MIMO relay network.

In this paper, we consider  an  MU-MIMO non-regenerative relay
network where each node is equipped with multiple antennas. We take
the effect of S-D link into the joint optimization of the source
precoding matrices and relay processing matrix, which is more
complicated than the relatively simple case without considering S-D
links~\cite{2010-YuanYu}. To our best knowledge, there is no such
work in the literature on the joint optimization of source precoding
and relay processing for MU-MIMO non-regenerative relay networks
with direct S-D links. Two major contributions of this paper over
the conventional works are as follows:
\begin{itemize}
\item We first introduce a general strategy to the joint design of  source precoding matrices
and relay processing matrix by transforming the network into a set
of parallel scalar sub-systems just as a point-to-point MIMO channel
under a relay modified  power constraint, and show that the capacity
based source precoding matrices and relay processing matrix
respectively share the same structures with the MSE based ones.
\item A nested iterative algorithm is presented  to solve  the
joint optimization of sources precoding and relay processing  based
on capacity and MSE respectively. Simulation results show that the
proposed algorithm outperforms the existing methods.
\end{itemize}

The rest of this paper is organized as follows. Section II
illustrates the system model. Section III presents the optimal
structures of source precoding and relay processing, and a nested
iterative algorithm to solve the joint optimization of sources
precoding and relay processing. Section IV devotes to the simulation
results. Finally, Section V concludes the paper.

\emph{Notations:} Lower-case letter, boldface lower-case letter, and
boldface upper-case letter denote scalar, vector, and matrix,
respectively. $\textsf{E}(\cdot)$, $\mathrm{tr}(\cdot)$,
$(\cdot)^{-1}$, $(\cdot)^{\dag}$, $|\cdot|$, and $\|\cdot\|_{F}$
denote expectation, trace, inverse,  conjugate transpose,
determinant, and Frobenius norm of a matrix, respectively.
$\mathbf{I}_{N}$ stands for  the identity matrix of order $N$.
$\mathrm{diag}(a_{1},\ldots,a_{N})$ is a diagonal matrix with the
$i$th diagonal entry $a_{i}$. $\log$ is of base $2$.
$\mathcal{C}^{M\times N}$ represents the set of $M\times N$ matrices
over complex field, and $\sim\mathcal{CN}(x,y)$ means satisfying  a
circularly symmetric complex Gaussian distribution with  mean $x$
and covariance $y$. $[x]^{+}$ denotes $\max\{0,x\}$.

\section{System Model}
\begin{figure}[!t]
\begin{center}
\includegraphics [width=3.0in]{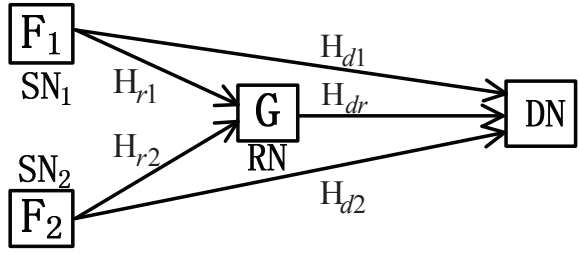}
\caption{The multiple-access relay network  with two source nodes, one relay node, and one destination node} \label{Fig-ModelUplink}
\end{center}
\end{figure}
We consider a  multiple access MIMO relay network with two source
nodes (SNs), one relay node~(RN) and one destination node~(DN) as
illustrated in Fig.~\ref{Fig-ModelUplink}, where the channel
matrices have been shown. The numbers of antennas equipped at the
SNs, RN and DN are $N_{s}, N_{r}$, and $N_{d}$, respectively. We
assume that there is only two SNs and both SNs have the same number
of antennas for simplicity. However, it is easy to be generalized to
the scenario of multiple SNs with different numbers of antennas at
each SN. In this paper, we consider a non-regenerative half-duplex
relaying strategy applied at the RN to process the received signals.
Thus, the transmission will take place in two phases. Suppose that
perfect synchronization has been established between SN$_{1}$ and
SN$_{2}$ prior to transmission, and both SN$_{1}$ and SN$_{2}$
transmit their independent  messages to the RN  and DN
simultaneously during the first phase. Then the RN processes the
received signals and forwards them to the DN during the second
phase.

Let $\mathbf{H}_{ri}\in \mathcal{C}^{N_{r}\times N_{s}},
\mathbf{H}_{di}\in \mathcal{C}^{N_{d}\times N_{s}}, $ and $
\mathbf{H}_{dr}\in \mathcal{C}^{N_{d}\times N_{r}} $ denote the
channel matrices of the $i$th SN to RN, to DN, and RN to DN,
respectively. Each entry of the channel matrices  is assumed to be
complex Gaussian variable with zero-mean and
variance~$\sigma^{2}_{h}$. Furthermore, all the channels involved
are assumed to be quasi-static i.i.d. Rayleigh fading combining with
large scale fading  over a common narrow-band. Let
$\mathbf{F}_{1}\in \mathcal{C}^{N_{s}\times N_{s}}$  and
$\mathbf{F}_{2}\in \mathcal{C}^{N_{s}\times N_{s}}$ denote the
precoding matrices for SN$_{1}$ and SN$_{2}$, respectively, which
satisfy the power constraint
$\textsf{E}[\mathbf{F}_{i}\mathbf{s}_{i}\mathbf{s}^{\dag}_{i}\mathbf{F}^{\dag}_{i}]=\mathrm{tr}(\mathbf{F}_{i}\mathbf{F}^{\dag}_{i})\leq
P_{i}$. Let $\mathbf{G}\in \mathcal{C}^{N_{r}\times N_{r}}$ denote
the relay processing  matrix.  Suppose that $\mathbf{n}_{r}\in
\mathcal{C}^{N_{r}\times 1}$ and $\mathbf{n}_{i}\in
\mathcal{C}^{N_{d}\times 1}$ are the noise vectors at RN and  DN,
respectively, and all noise are independent and identically
distributed  additive white Gaussian noise (AWGN) with zero-mean and
unit variance. Then, the  baseband signal vectors $\mathbf{y}_{1}$
and $\mathbf{y}_{2}$ received at the DN during the two consecutive
phases can be expressed as follows:
%
\begin{IEEEeqnarray} {rCl} \label{eq:model}
\underbrace{
\left[
  \begin{array}{c}
    \mathbf{y}_{1} \\
    \mathbf{y}_{2} \\
  \end{array}
\right]}_{\mathbf{Y}} &=&
\underbrace{
\left[
  \begin{array}{c}
    \mathbf{H}_{d1} \\
    \mathbf{H}_{dr}\mathbf{G}\mathbf{H}_{r1}
  \end{array}
\right]
}_{\mathbf{H}_{1}} \mathbf{F}_{1}\mathbf{s}_{1} +
\underbrace{
\left[
  \begin{array}{c}
 \mathbf{H}_{d2} \\
 \mathbf{H}_{dr}\mathbf{G}\mathbf{H}_{r2}\\
  \end{array}
\right]
}_{\mathbf{H}_{2}}
\mathbf{F}_{2}\mathbf{s}_{2}
+ \nonumber\\&&
\underbrace{
\left[
  \begin{array}{ccc}
    \mathbf{I}_{N_d} & \mathbf{0}& \mathbf{0} \\
    \mathbf{0}&\mathbf{H}_{dr}\mathbf{G} & \mathbf{I}_{N_d} \\
  \end{array}
\right]}_{\mathbf{H}_{3}}
\underbrace{
\left[
  \begin{array}{c}
    \mathbf{n}_{1} \\
    \mathbf{n}_{r}  \\
    \mathbf{n}_{2} \\
  \end{array}
\right]}_{\mathbf{N}},
\end{IEEEeqnarray}
%
%
where $\mathbf{s}_{i}\in \mathcal{C}^{N_{s}\times 1}$ is assumed to
be a zero-mean circularly symmetric complex Gaussian signal vector
transmitted by the $i$th SN and satisfies
$\textsf{E}(\mathbf{s}_{i}\mathbf{s}^{\dag}_{i})=\mathbf{I}_{N_{s}}$.
Let $\textbf{Y}$, $\textbf{H}_i$ ($i=1,2,3$), and $\textbf{N}$,
shown in  (\ref{eq:model}), denote the effective receive signal,
effective channels and effective noise respectively. Then
$\mathbf{H}_{3}\textsf{E}[\mathbf{NN}^{\dag}]\mathbf{H}_{3}^{\dag} =
\mathbf{H}_{3}\mathbf{H}_{3}^{\dag}=
\mathrm{diag}(\mathbf{I}_{N_{d}},\mathbf{R})$, where
$\mathbf{R}=\mathbf{I}_{N_{d}}+
\mathbf{H}_{dr}\mathbf{GG}^{\dag}\mathbf{H}^{\dag}_{dr}$ is the
covariance matrix of the effective noise  at the DN during the
second phase.
\section{Optimal coordinates of Joint Source and Relay  Design }
In this section, the capacity and MSE  for the MMSE detector with
successive interference cancelation (SIC) at DN  are analyzed. Then,
we will  exploit the optimal structures of source precoding and
relay processing based on capacity and MSE respectively. Then a new
algorithm on how to jointly optimize the sources precoding matrices
and the relay processing matrix is proposed to maximize the capacity
or minimize MSE of the entire network.

\subsection{Decoding Scheme}
Conventional receivers such as matched filter (MF), zero-forcing
(ZF), and MMSE decoder have been well studied in the previous works.
The MF receiver has bad performance in the high SNR region, whereas
the ZF produces a noise enhancement effect in the low SNR region.
The MMSE detector with SIC has significant advantage over MF and ZF,
which is information lossless  and
optimal~\cite{2005-DavidTse-book}. Therefore, we consider the
MMSE-SIC receiver at the DN and first decode the signal from SN$_2$
without loss of generality. With the predetermined decoding order,
the interference from SN$_{2}$ to SN$_{1}$ is virtually absent. To
exploit the optimal structures of the matrices at the SNs, we first
set up the  RN  with a fixed processing  matrix $\mathbf{G}$ without
considering the power control. With the predetermined decoding
order, the MMSE receive filter for SN$_i$ ($i=1,2$) is given
as~\cite{1993-EstimationTheory}\cite{2008-Soren}:
\begin{IEEEeqnarray}{rCl}
\mathbf{A}^{\mathrm{MMSE}}_{i}=\mathbf{F}^{\dag}_{i}\mathbf{H}^{\dag}_{i}(\mathbf{H}_{i}\mathbf{F}_{i}\mathbf{F}^{\dag}_{i}\mathbf{H}^{\dag}_{i}+\mathbf{R}_{Z_{i}})^{-1},
\end{IEEEeqnarray}
where
$
\mathbf{R}_{Z_{1}}\triangleq \mathbf{H}_{3}\mathbf{H}^{\dag}_{3}~\mathrm{and}~\mathbf{R}_{Z_{2}}\triangleq \mathbf{H}_{3}\mathbf{H}^{\dag}_{3}+\mathbf{H}_{1}\mathbf{F}_{1}\mathbf{F}^{\dag}_{1}\mathbf{H}^{\dag}_{1}.
$
Then, the MSE-matrix for SN$_{i}$ can be expressed as:
\begin{IEEEeqnarray}{rCl}\label{eq:MMSE-i}
\mathbf{E}_{i}&=&\textsf{E}\left[(\mathbf{A}^{\mathrm{MMSE}}_{i}\mathbf{Y}_{i}-\mathbf{s}_{i})(\mathbf{A}^{\mathrm{MMSE}}_{i}\mathbf{Y}_{i}-\mathbf{s}_{i})^{\dag} \right] \nonumber\\
&=&
\left( \mathbf{I}_{N_{s}}+\mathbf{F}^{\dag}_{i}\mathbf{H}^{\dag}_{i} \mathbf{R}^{-1}_{Z_{i}}\mathbf{H}_{i}\mathbf{F}_{i}\right)^{-1} ,
\end{IEEEeqnarray}
where $\mathbf{Y}_{1}=\mathbf{Y}-\mathbf{H}_{2}\mathbf{F}_{2}\mathbf{s}_{2}$ and $\mathbf{Y}_{2}=\mathbf{Y}$. Hence, the capacity for SN$_{i}$ is given as~\cite{2005-DavidTse-book}
\begin{IEEEeqnarray}{rCl}\label{eq:Rate-Ri}
{C}_{i}=\log\left|\mathbf{I}_{N_{s}}+\mathbf{F}^{\dag}_{i}\mathbf{H}^{\dag}_{i} \mathbf{R}^{-1}_{Z_{i}}\mathbf{H}_{i}\mathbf{F}_{i} \right|=\log \left|\mathbf{E}^{-1}_{i} \right|.
\end{IEEEeqnarray}

\subsection{Optimal Precoding  Matrices at SNs} \label{sec:OPMatS}
In this subsection,  we will   introduce two  lemmas, which will be
used to exploit the optimal source  precoding matrices  and  relay
processing matrix, respectively.
\newtheorem{lemma}{Lemma}
\begin{lemma}\label{Lemma:A}
For a matrix $\mathbf{A}$, if matrix $\mathbf{B}$ is a positive
definite matrix, and
$\mathbf{C}=\mathbf{A}\mathbf{B}^{-1}\mathbf{A}^{\dag}$, then
$\mathbf{C}$  is an Hermitian and positive semidefinite matrix
(HPSDM).
\end{lemma}
\begin{IEEEproof}
Since $\mathbf{B}$ is a positive definite matrix, then
$\mathbf{B}^{-1}$ is also a positive definite matrix. For any
non-zero column vector $\textbf{\textit{x}}$, let
$\textbf{\textit{y}}=\mathbf{A}^{\dag}\textbf{\textit{x}}$. Then we
have $\textbf{\textit{x}}^{\dag}\mathbf{C}\textbf{\textit{x}}
=\textbf{\textit{x}}^{\dag}\mathbf{A}\mathbf{B}^{-1}\mathbf{A}^{\dag}\textbf{\textit{x}}
=\textbf{\textit{y}}^{\dag}\mathbf{B}^{-1}\textbf{\textit{y}}\geq
0$, which implies that $\mathbf{C}$ is an HPSDM.
\end{IEEEproof}
\begin{lemma}\label{Lemma:B}
If $\mathbf{A}$ and $\mathbf{B}$ are positive semidefinite matrices,
then, $0\leq \mathrm{tr}(\mathbf{AB})\leq
\mathrm{tr}(\mathbf{A})\mathrm{tr}(\mathbf{B})$, and, there is an
$\alpha \in[0,1]$, such that $\mathrm{tr}(\mathbf{AB})=\alpha
\mathrm{tr}(\mathbf{A})\mathrm{tr}(\mathbf{B})$.
\end{lemma}
\begin{IEEEproof}
See~\cite[page 269]{1976-Mathematical}.
\end{IEEEproof}

Since $\mathbf{R}_{Z_{i}}~(i=1,2)$  is positive definite
matrix~\cite{1996-Statistical}, according  to
\emph{Lemma~\ref{Lemma:A}},
$\mathbf{H}_{si}=\mathbf{H}^{\dag}_{i}\mathbf{R}^{-1}_{Z_{i}}\mathbf{H}_{i}$
is HPSDM, which can be decomposed as:
\begin{IEEEeqnarray}{rCl}
\mathbf{H}_{si}=\mathbf{U}_{i}\mathbf{\Lambda}_{i}\mathbf{U}^{\dag}_{i},
\end{IEEEeqnarray}
with unitary matrix  $\mathbf{U}_{i}$, and non-negative diagonal
matrices $\mathbf{\Lambda}_{i}$, which diagonal entries are in
descending order. One of our main results of this paper is as below.

\newtheorem{Theorem}{Propositon}
\begin{Theorem}\label{Lemma:Pro}
For  a given  matrix\footnote{The relay power constraint problem    will be deal with directly by an iterative algorithm later.} $\mathbf{G}$ and predetermined decoding order,
the precoding matrix for SN$_i$ with the following canonical  form
 \begin{equation}\label{Eq:optimal-source}
 \mathbf{F}_{i}=\mathbf{U}_{i}\mathbf{\Sigma}_{i}~~(i=1,2)
 \end{equation}
is optimal with the water-filling power allocation policy~(Policy-A)
based on capacity or with the inverse water-filling power allocation
policy~(Policy-B) based on MSE, where:
\begin{IEEEeqnarray}{rCl}
\mathbf{\Sigma}^{2}_{i}&=&\left[\mu-\mathbf{\Lambda}^{-1}_{i}\right]^{+}~~~~~~~~~ (\mathrm{Policy-A}),~~\IEEEyessubnumber\\
\mathbf{\Sigma}^{2}_{i}&=&\left[\mu\mathbf{\Lambda}^{-1/2}_{i}-\mathbf{\Lambda}^{-1}_{i}\right]^{+}~  (\mathrm{Policy-B}),\IEEEyessubnumber\\
\mathrm{s.t}&:& \mathrm{tr}(\mathbf{\Sigma}^{2}_{i})=P_i. \IEEEyessubnumber
\end{IEEEeqnarray}
\end{Theorem}
\begin{IEEEproof}
Substituting $\mathbf{F}_{1}$ in (\ref{Eq:optimal-source}) into
(\ref{eq:Rate-Ri}) and (\ref{eq:MMSE-i}), we  respectively have:
\begin{IEEEeqnarray}{rCl}
{C}_{1}&=&
\log\left|\mathbf{I}_{N_{s}}+\mathbf{\Sigma}^{2}_{1}\mathbf{\Lambda}_{1} \right|,
\nonumber\\
\mathrm{tr}(\mathbf{E}_{1})
&=&\mathrm{tr}\left\{( \mathbf{I}_{N_{s}}+\mathbf{\Sigma}^{2}_{1}\mathbf{\Lambda}_{1})^{-1}\right\}.\nonumber
\end{IEEEeqnarray}
According to KKT conditions~\cite{2004-Boyd-CO}, the Policy-A and
Policy-B can make the capacity $C_1$ maximized and the MSE
$\mathrm{tr}(\mathbf{E}_{1})$  minimized, respectively,  under the
power control $P_1$ at SN$_1$. This implies that $\mathbf{F}_1$ is
optimal. After deciding $\mathbf{F}_1$, and substituting the
$\mathbf{F}_1$ into $\mathbf{R}_{Z_{2}}$, we can prove that
$\mathbf{F}_2$ is optimal.
\end{IEEEproof}

\subsection{A Nearly Optimal Processing Matrix at Relay} \label{sec:matrixG}

In this subsection, we first exploit the structure of relay
processing matrix based on capacity for given $\mathbf{F}_1$ and
$\mathbf{F}_2$. Then, we show that the same structure matrix at RN
can make the MSE of the entire network near to minimum with a
different power allocation policy.
The capacity of the entire network is~\cite{2005-DavidTse-book}
\begin{IEEEeqnarray}{rCl}\label{eq:sum-rate}
&&C=
\log \left|\mathbf{H}_{1}\mathbf{\Pi}_{1}\mathbf{H}^{\dag}_{1}+
\mathbf{H}_{2}\mathbf{\Pi}_{2}\mathbf{H}^{\dag}_{2}+
\mathbf{H}_{3}\mathbf{H}^{\dag}_{3}
\right|- \log\left|\mathbf{H}_{3}\mathbf{H}^{\dag}_{3}\right|,\nonumber
\end{IEEEeqnarray}
where $\mathbf{\Pi}_{i}=\mathbf{F}_{i}\mathbf{F}^{\dag}_{i}$.
According to the determinant  expansion  formula of the block
matrix~\cite{1991-Roger-MA}, (\ref{eq:sum-rate}) can be rewritten
as:
\begin{IEEEeqnarray}{rCl}\label{eq:Sum-2}
C&=&\log \left|\mathbf{T} \right|+\log \left|\mathbf{H}_{dr}\mathbf{G}\mathbf{K}\mathbf{G}^{\dag}\mathbf{H}^{\dag}_{dr} + \mathbf{R} \right| -\log\left|\mathbf{R}\right|,
\end{IEEEeqnarray}
where
\begin{IEEEeqnarray}{rCl}
\mathbf{T}&=&\mathbf{I}_{N_{d}}+\sum_{i=1}^{2}\mathbf{H}_{di}\mathbf{\Pi}_{i}\mathbf{H}^{\dag}_{di},\IEEEyessubnumber\\
\mathbf{K}&=&\sum_{i=1}^{2}\mathbf{H}_{ri}\mathbf{\Pi}_{i}\mathbf{H}^{\dag}_{ri}-
\mathbf{\widetilde{K}},\IEEEyessubnumber \label{eq:K}\\
\mathbf{\widetilde{K}}&=&\left(
\sum_{i=1}^{2}\mathbf{H}_{ri}\mathbf{\Pi}_{i}\mathbf{H}^{\dag}_{di}\right)
\mathbf{T}^{-1} \left(\sum_{i=1}^{2}
\mathbf{H}_{di}\mathbf{\Pi}_{i}\mathbf{H}^{\dag}_{ri} \right) \IEEEyessubnumber \label{eq:KK}.
\end{IEEEeqnarray}

Let $\Delta=\log|\mathbf{T}|$, which is independent of $\mathbf{G}$.
Then, for given  $\mathbf{F}_{1}$ and $\mathbf{F}_{2}$, the problem
on  maximum   capacity of the network  can be formulated as
\begin{IEEEeqnarray}{rCl} \label{eq:cost-F}
\arg&&
  \max_{\mathbf{G} }~ C= \log\left|\mathbf{H}_{dr}\mathbf{G}\mathbf{K}\mathbf{G}^{\dag}\mathbf{H}^{\dag}_{dr} + \mathbf{R}\right|-  \log\left| \mathbf{R} \right|,   \IEEEyessubnumber    \label{eq:MaxC-G}\\
\mathrm{s.t.}&&~ \mathrm{tr}\left\{\mathbf{G}\left(\mathbf{I}_{N_{r}}+\sum^{2}_{i=1}\mathbf{H}_{ri}\mathbf{\Pi}_{i}\mathbf{H}^{\dag}_{ri}\right)\mathbf{G}^{\dag} \right\} \leq  P_{r}   \IEEEyessubnumber \label{eq:PowerR-G}.
\end{IEEEeqnarray}
To solve this problem, and find a nearly  optimal processing matrix
$\mathbf{G}$,   due to $\mathbf{K}=\mathbf{K}^{\dag}$, we first
decompose $\mathbf{K}$  based on eigenvalue decomposition, and  then
decompose  $\mathbf{H}_{dr}$ based on singular value decomposition,
i.e.,
\begin{IEEEeqnarray}{rCl}
\mathbf{K}&=&\mathbf{U}_{K}\mathbf{\Lambda}_{K}\mathbf{U}^{\dag}_{K},\nonumber\\
\mathbf{H}_{dr}&=&\mathbf{U}_{H}\mathbf{\Theta}\mathbf{V}^{\dag}_{H},\nonumber
\end{IEEEeqnarray}
where $\mathbf{U}_{K},\mathbf{U}_{H}$  and $\mathbf{V}_{H}$ are
unitary matrices, and
$\mathbf{\Lambda}_{K}=\mathrm{diag}(\lambda_{1},\cdots,\lambda_{N_{r}})$
is  an $N_{r}\times N_{r}$ diagonal matrix, and
$\mathbf{\Theta}=\mathrm{diag}(\theta_{1},\cdots,\theta_{r})$ is an
$N_{r}\times N_{r}$ diagonal matrix, which diagonal entries are in
descending order.

From (\ref{eq:MaxC-G}), it is easy to verify that the optimal  left
canonical of $\mathbf{G}$ is still given by
$\mathbf{V}_{H}$~\cite{2007-Xiaojun}. But, it is intractable to find
the optimal right canonical for the processing matrix $\mathbf{G}$,
because there is no matrix which can achieve the diagonalization of
both the capacity cost function~(\ref{eq:MaxC-G}) and the power
constraint~(\ref{eq:PowerR-G}). But, we can modify the power
constraint (\ref{eq:PowerR-G}) to another expression to find a
matrix which has the desired property. Due to $\mathbf{K}$
is a deterministic matrix for the fixed sources precoding matrices, (\ref{eq:PowerR-G}) can be rewritten as
\begin{multline}
\mathrm{tr}\{\mathbf{G}(\mathbf{I}_{N_{r}}+\mathbf{K})\mathbf{G}^{\dag}\}+
\mathrm{tr}\{\mathbf{\widetilde{K}}\mathbf{G}^{\dag}\mathbf{G}\}
=\\
\mathrm{tr}\left\{\mathbf{G}\left(\mathbf{I}_{N_{r}}+\sum^{2}_{i=1}\mathbf{H}_{ri}\mathbf{\Pi}_{i}\mathbf{H}^{\dag}_{ri}\right)\mathbf{G}^{\dag}\right\}
\leq P_{r}.\nonumber
\end{multline}
Since  $\mathbf{T}$ is a positive definite matrix,
according to \emph{Lemma~\ref{Lemma:A}}, $\mathbf{\widetilde{K}}$ in (\ref{eq:KK}) is also a  positive semidefinite matrix.
According to \emph{Lemma~\ref{Lemma:B}},
the new power constraint at the RN can be expressed as
\begin{multline} \label{eq:NewPowerConstr}
\mathrm{tr}\{\mathbf{G}(\mathbf{I}_{N_{r}}+\mathbf{K})\mathbf{G}^{\dag}\}+
\alpha\mathrm{tr}\{\mathbf{\widetilde{K}}\}\mathrm{tr}\{\mathbf{G}^{\dag}\mathbf{G}\}
\approx \\
\mathrm{tr}\left\{\mathbf{G}(\mathbf{I}_{N_{r}}+\sum^{2}_{i=1}\mathbf{H}_{ri}\mathbf{F}_{i}\mathbf{F}^{\dag}_{i}\mathbf{H}^{\dag}_{ri})\mathbf{G}^{\dag}\right\} \leq P_{r},
\end{multline}
where the exact value $\alpha$ can be found by an iterative method. Thus, applying the results in~\cite{2007-Xiaojun}\cite{2010-YuanYu}, the processing matrix $\mathbf{G}$ with the following structure  can achieve the desired diagonalization for both capacity  cost function (\ref{eq:MaxC-G}) and the new power constraint (\ref{eq:NewPowerConstr}), and will be optimal~\cite{2007-Xiaojun}:
\begin{equation}\label{eq:optimal-G}
\mathbf{G}=\mathbf{V}_{H}\mathbf{\Xi} \mathbf{U}^{\dag}_{K},
\end{equation}
where $\mathbf{\Xi}^{2}=\mathrm{diag}(\xi_{1},\cdots,\xi_{N_r})$ can be solved by optimization method~\cite{2007-Xiaojun}.

Let $\kappa=\mathrm{tr}\{\mathbf{\widetilde{K}}\}$. Substituting $\mathbf{G}$ into (\ref{eq:MaxC-G}), and using the new power constraint (\ref{eq:NewPowerConstr}) to replace (\ref{eq:PowerR-G}), the problem (\ref{eq:cost-F}) to find $\xi_{i}$ becomes
\begin{IEEEeqnarray}{rCl}
\arg&&
  \max_{\xi_{1},~\ldots,~\xi_{N_{r}}}~ C(\xi_{i})=\sum^{N_{r}}_{i=1}\log\frac{\theta^{2}_{i}\xi_{i}\lambda_{i}+\theta^{2}_{i}\xi_{i}+1 }{\theta^{2}_{i}\xi_{i}+1 }, \IEEEyessubnumber       \label{eq:NewMaxC-G}\\
\mathrm{s.t.}&&~ \sum^{N_{r}}_{i=1}(\lambda_{i}+\alpha \kappa+1)\xi_{i} \leq  P_{r}~~ \mathrm{and} ~~\xi_{i}\geq 0, ~ \forall i~. \IEEEyessubnumber  \label{eq:NewPower-R}
\end{IEEEeqnarray}
Then,  this optimization problem with respect to $\xi_{i}$ is similar to a problem solved in~\cite{2007-Xiaojun,2010-YuanYu}. Then we have
\begin{IEEEeqnarray}{rCl}
&&\xi_{i}=\frac{1}{2\theta^{2}_{i}(\lambda_{i}+1)} \left[\sqrt{\lambda^{2}_{i}+\frac{4\lambda_{i}\theta^{2}_{i}(\lambda_{i}+1)\mu}{\lambda_{i}+1+\alpha \kappa}}-\lambda_{i}-2\right]^{+}\label{eq:xi} \\
&&\sum^{N_{r}}_{i=1}(\lambda_{i}+1+\alpha \kappa)\xi_{i}\leq P_{r}. \label{eq:mu}
\end{IEEEeqnarray}
where $\mu$ in (\ref{eq:xi})  is decided by (\ref{eq:mu}).

Next, we will show that the same structure matrix $\mathbf{G}$ can
also  make the  MSE of the entire network near to minimum with a
different power allocation matrix $\mathbf{\Xi}$ for given
$\mathbf{F}_1$ and $\mathbf{F}_2$. Due to the total MSE can be
expressed as:
\begin{IEEEeqnarray}{rCl}
J(\mathbf{G})&=&\mathrm{tr}(\mathbf{E}_{1})+\mathrm{tr}(\mathbf{E}_{2})\nonumber\\
&\overset{a}{\leq}&\mathrm{tr}(\mathbf{\tilde{E}}_{1})+\mathrm{tr}(\mathbf{E}_{2}) \nonumber\\
&=&
\mathrm{tr}\left\{( \mathbf{I}_{2N_{d}}+\mathbf{F}^{\dag}\mathbf{H}^{\dag} \mathbf{R}^{-1}_{Z_{1}}\mathbf{H}\mathbf{F})^{-1}\right\} \nonumber\\
&\overset{b}{=}&
\mathrm{tr}(\mathbf{I}_{2N_{d}})-\mathrm{tr}\left\{
(\mathbf{R}_{Z_{1}}+\mathbf{H}\mathbf{F}\mathbf{F}^{\dag}\mathbf{H}^{\dag})^{-1}\mathbf{H}\mathbf{F}\mathbf{F}^{\dag}\mathbf{H}^{\dag} \right\}
\nonumber\\
&=&
\mathrm{tr}\left\{(\mathbf{R}_{Z_{1}}+\mathbf{H}\mathbf{F}\mathbf{F}^{\dag}\mathbf{H}^{\dag})^{-1}\mathbf{R}_{Z_{1}} \right\}\nonumber \\
&\overset{c}{=}&
\beta\mathrm{tr}\left\{(\mathbf{H}_{dr}\mathbf{G}\mathbf{K}\mathbf{G}^{\dag}\mathbf{H}^{\dag}_{dr}
+ \mathbf{R})^{-1} \right\} \mathrm{tr}\left\{(\mathbf{I}_{N_{d}}+
\mathbf{R})\right\} \nonumber\\
&\triangleq&
\beta\tilde{J}(\mathbf{G}),
\label{eq:J-MMSE}
\end{IEEEeqnarray}
where $\mathbf{F}=\mathrm{diag}(\mathbf{F}_1,\mathbf{F}_2)$,
$\mathbf{H}=[\mathbf{H}_{1}~ \mathbf{H}_2]$, $\beta$ is a scalar
factor. In (\ref{eq:J-MMSE}), (a) come from the fact that noise is
enhanced by using
$\mathbf{\tilde{R}}_{Z_1}=\mathbf{H}_{3}\mathbf{H}^{\dag}_{3}+\mathbf{H}_{2}\mathbf{\Pi}_{2}\mathbf{H}^{\dag}_{2}$
to replace $\mathbf{R}_{Z_1}$ in calculating
$\mathrm{tr}(\mathbf{\tilde{E}}_{1})$, (b) follows from Woodbury
identity and $\mathrm{tr}(\mathbf{AB})=\mathrm{tr}(\mathbf{BA})$,
and (c) follows from \emph{Lemma~\ref{Lemma:B}} and Schur complement
to inverse a block matrix~\cite{1991-Roger-MA}. From
(\ref{eq:J-MMSE}), to minimize the  $J(\mathbf{G})$ is equivalent to
minimize $\tilde{J}(\mathbf{G})$.
Then, for given  $\mathbf{F}_{1}$ and $\mathbf{F}_{2}$, the optimal
$\mathbf{G}$  to minimize MSE is
\begin{IEEEeqnarray}{rCl} \label{eq:G-MSE}
&\arg&~
  \min_{\mathbf{G} }~\tilde{J}(\mathbf{G} ),\IEEEyessubnumber \\
  &\mathrm{s.t.}&:~~ ~~  (\ref{eq:NewPowerConstr}).  \IEEEyessubnumber
\end{IEEEeqnarray}
From the analysis above, the structure  of  $\mathbf{G}$ in
(\ref{eq:optimal-G})   can also achieve the diagonalization of the
equation (\ref{eq:G-MSE}), but, has a new power allocation matrix
$\mathbf{\Xi}$ different from that of capacity based one. Then,
substituting $\mathbf{G}$ in (\ref{eq:optimal-G}) into
(\ref{eq:G-MSE}) to find the new $\mathbf{\Xi}$,  (\ref{eq:G-MSE})
becomes
\begin{IEEEeqnarray}{rCl}
&\arg&
\min_{\xi_{1},\ldots,~\xi_{N_{r}}} \tilde{J}(\xi_{i}),\IEEEyessubnumber\\
&\mathrm{s.t.}&: ~~(\ref{eq:NewPower-R}) \IEEEyessubnumber.
\end{IEEEeqnarray}
where
\begin{IEEEeqnarray}{rCl}
\tilde{J}(\xi_{i})=\left(\sum\limits^{N_{r}}_{i=1}\left(\theta^{2}_{i}\lambda_{i}\xi_{i}+\theta^{2}_{i}\xi_{i}+1 \right )^{-1} \right)\left(\sum\limits^{N_{r}}_{i=1}(\theta^{2}_{i}\xi_{i}+2)\right).\nonumber
\end{IEEEeqnarray}
This problem  can be solved by numerical optimization
methods~\cite{2004-Boyd-CO}.

\subsection{Iterative Algorithm}
In the above discussion, with predetermined decoding order and fixed
$\mathbf{G}$, $\mathbf{F}_{1}$ and $\mathbf{F}_{2}$  can be
optimized; For $\mathbf{F}_{1}$ and $\mathbf{F}_{2}$, $\mathbf{G}$
can be optimized.  Therefore, we propose an iterative algorithm  to
jointly optimize $\mathbf{F}_{1},\mathbf{F}_{2}$ and $\mathbf{G}$
based on capacity. Note that, the MSE based algorithm can be easily
obtained as well. The convergence analysis of the proposed iterative
algorithm is intractable. But, it can yield much better performance
than the existing methods, which will be demonstrated by the
simulation results in the next section.

In summary, we outline the nested iterative algorithm as follows:
\begin{algorithm}
\caption{\textbf{:}~A nested iterative algorithm.}
 \begin{algorithmic}[]
 \STATE ${\bullet}$
\textbf{Initialization:} $\mathbf{G}$.
 \STATE ${\bullet}$ {\textbf{Repeat:}} Update $k:=k+1$;
 \STATE $\mathbf{~~}$ -- Compute $\mathbf{F}^{(k)}_{1}$ based on $\mathbf{G}^{(k)}$;
 \STATE $\mathbf{~~}$ -- Compute $\mathbf{F}^{(k)}_{2}$ based on $\mathbf{G}^{(k)}$ and  $\mathbf{F}^{(k)}_{1}$;
\STATE $\mathbf{~~}$ -- Compute $\mathbf{G}^{(k+1)}=\mathbf{V}_{H}\mathbf{\Xi}\mathbf{U}_{K}$ based on $\mathbf{F}^{(k)}_{1}$ and $\mathbf{F}^{(k)}_{2}$ by the following inner repeat to find $\mathbf{\Xi}$;\\
\begin{description}
 \STATE${\circ}$~~\textbf{Initial:}  $\alpha $;\\
 \STATE${\circ}$~~\textbf{Inner Repeat :} Update $n:=n+1$;
 \STATE${~~}$~~-- Compute $\mathbf{\Xi}^{(n)}$ based on $\alpha^{(n)}$;
 \STATE${~~}$~~-- Compute $\alpha^{(n+1)}$ based on $\mathbf{\Xi}^{(n)}$;
 \STATE${\circ}$ \textbf{Inner Until:} Convergence.
 \end{description}
 \STATE $\bullet$ \textbf{Until:} The termination  criterion is satisfied.
\end{algorithmic}
\end{algorithm}

\section{Simulation Results}

\begin{figure}[!t]
\begin{center}
\includegraphics [width=3.5in]{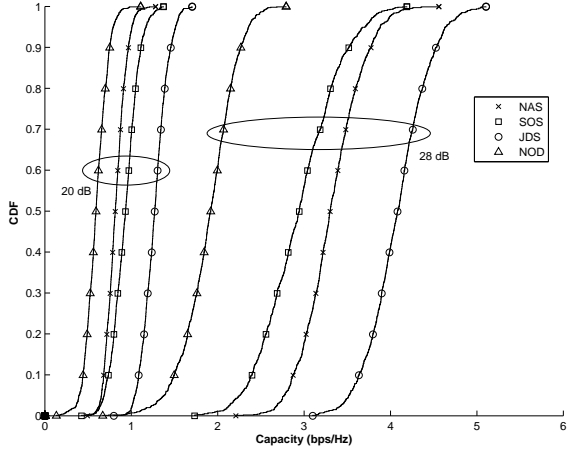}
\caption{CDF of the capacity for different power constraints, $P_1=P_2=P_r=20\mathrm{dB}$ and~ $P_1=P_2=P_r=28\mathrm{dB}$, $N_s=N_r=N_d=4$, $\ell_{sd}=10,~\ell_{sr}=\ell_{rd}=5$} \label{Fig-CDF}
\end{center}
\end{figure}
\begin{figure}[!t]
\begin{center}
\includegraphics [width=3.5in]{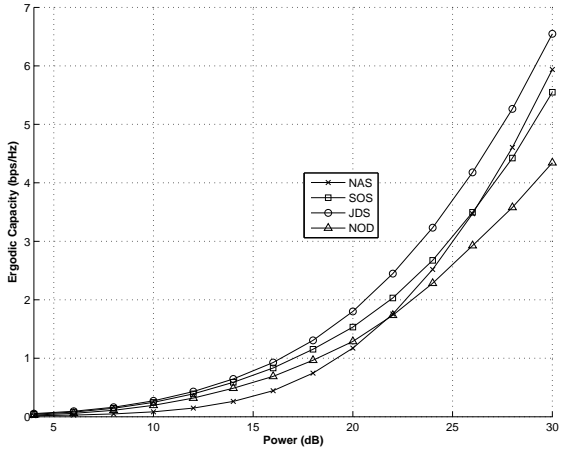}
\caption{The capacity versus the power constraints $P_i~ (i=1,2,r)$ (dB), $P_1=P_2=P_r$, and $N_s=N_r=N_d=4$, $\ell_{sd}=10,~\ell_{sr}=\ell_{rd}=5$.} \label{Fig-CvsP}
\end{center}
\end{figure}
\begin{figure}[!t]
\begin{center}
\includegraphics [width=3.5in]{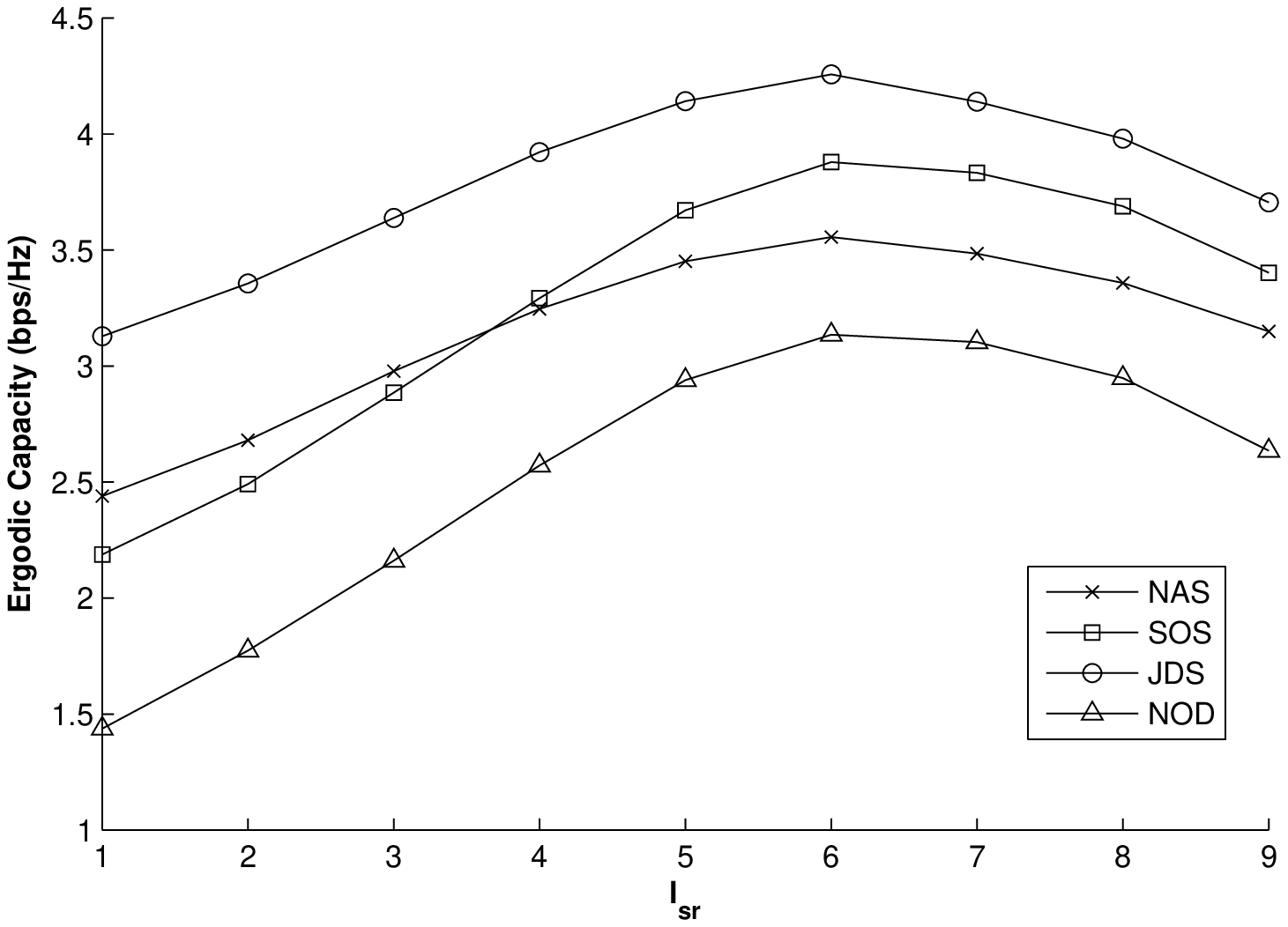}
\caption{The capacity versus the distance between source-to-relay~($\ell_{sr})$,~ $\ell_{sd}=10,~\ell_{rd}=\ell_{sd}-\ell_{sr}$, and $P_1=P_2=P_r=26\mathrm{dB}$, $N_s=N_r=N_d=4$.} \label{Fig-Lsr}
\end{center}
\end{figure}
\begin{figure}[!t]
\begin{center}
\includegraphics [width=3.5in]{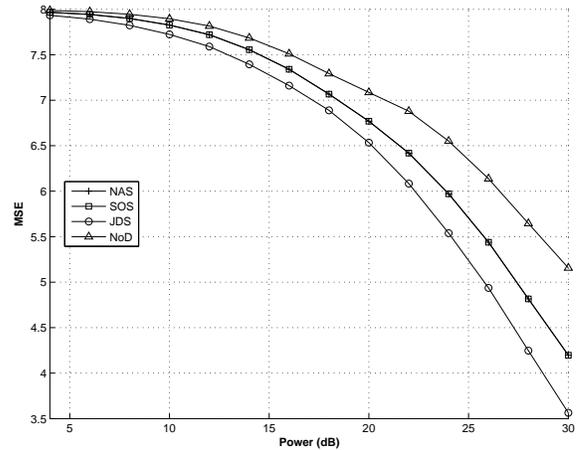}
\caption{The sum-MSE versus the power constraints $P_i~ (i=1,2,r)$ (dB), $P_1=P_2=P_r$, and $N_s=N_r=N_d=4$, $\ell_{sd}=10,~\ell_{sr}=\ell_{rd}=5$.} \label{Fig-MSEvsPower}
\end{center}
\end{figure}

\begin{figure}[!t]
\begin{center}
\includegraphics [width=3.5in]{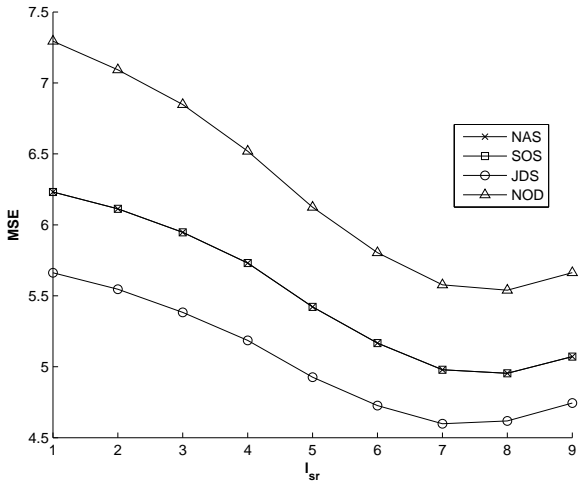}
\caption{The sum-MSE versus the distance between source-to-relay~($\ell_{sr})$,~ $\ell_{sd}=10,~\ell_{rd}=\ell_{sd}-\ell_{sr}$, and $P_1=P_2=P_r=26\mathrm{dB}$, $N_s=N_r=N_d=4$.} \label{Fig-MSELsr}
\end{center}
\end{figure}

In this section, simulation results are carried out to verify the performance superiority of the proposed joint source-relay design scheme~(JDS) for MU-MIMO relay network with direct links.  We first compare the proposed scheme with other three schemes in terms of  the
ergodic capacity  and the Cumulative Distribution Function~(CDF)
of instantaneous capacity of the MIMO relaying  networks, and then compare the sum-MSE of the networks. The alternative schemes are:
\begin{enumerate}
\item[(1)]\emph{ Naive scheme~(NAS)}:  The source covariances are fixed to be  scaled by the identity matrices $\frac{P_{1}}{N_{S}}\mathbf{I}$ and $\frac{P_{2}}{N_{S}}\mathbf{I}$ at SN$_{1}$ and SN$_{2}$, respectively, and the relay  processing  matrix  is $\mathbf{G}=\eta \mathbf{I}$, where $\eta=\sqrt{\frac{P_{r}}{\mathrm{tr}(\mathbf{I}+\sum^{2}_{i=1}\mathbf{H}_{ri}\mathbf{F}_{i}\mathbf{F}^{\dag}_{i}\mathbf{H}^{\dag}_{ri})}}$ is a power control factor. The S-D links contribution is included.
\item[(2)] \emph{Suboptimal scheme~(SOS)}: This scheme is proposed in~\cite{2010-YuanYu} for MU-MIMO relay network without considering S-D links in design. But, the S-D links contribution of capacity is included in the simulation for fair comparison. Note that this scheme is optimal for the scenario without considering  the S-D links.
\item[(3)] \emph{No-direct links scheme~(NOD)}: This scheme is like SOS, but, without S-D links contribution.
\end{enumerate}
Noting that both SOS and NOD have different power control polices to
accommodate the capacity and MSE criterions. In the simulations, we
consider a linear two-dimensional symmetric network geometry  as
depicted in Fig.~\ref{Fig-ModelUplink}, where both SNs are deployed
at the same position, and the distance between SNs (or RN) and DN is
set to be $\ell_{sd}$ (or $\ell_{rd}$), and
$\ell_{sd}=\ell_{sr}+\ell_{rd}$.   The channel gains are modeled as
the combination of large scale fading~(related to distance) and
small scale fading~(Rayleigh fading), and all channel matrices have
i.i.d. $\mathcal{CN}(0,\frac{1}{\ell^{\tau}})$ entries, where $\ell$
is the distance between two nodes, and $\tau=3$ is the path loss
exponent.

Fig.~\ref{Fig-CDF}-\ref{Fig-Lsr} are based on capacity criterion.
Fig.~\ref{Fig-CDF} shows the CDF of instantaneous capacity for different power constraints, when all nodes positions are fixed.  Fig.~\ref{Fig-CvsP} shows the capacity of the network versus the power constraints, when all nodes positions are fixed. These two figures show that capacity offered by the proposed relaying scheme is better than both SOS and NOD schemes at all SNR regime, especially at high SNR regime. The naive scheme surpasses  both SOS and NOD schemes at high SNR regime, which demonstrates that the direct S-D link should not be ignored in design.
Fig.~\ref{Fig-Lsr} shows the capacity of the network versus the distance ($\ell_{sr}$) between SNs and RN, for fixed $\ell_{sd}$. It is clear that the capacity offered by the proposed scheme
is better than those by the SOS, NAS and NOD schemes. NOD scheme is  the worst performance scheme at any relay position at moderate and high SNR regimes.

Fig.~\ref{Fig-MSEvsPower} and Fig.~\ref{Fig-MSELsr}, are based on   MSE criterion, the similar  conclusions can be drawn.
\section{Conclusion}

In this paper, we propose a optimal structure of the source
precoding matrices and relay processing matrix for  MU-MIMO
non-regenerative relay network  with  direct S-D links based on
capacity and  MSE respectively. We show that the capacity based
optimal source precoding matrices share the same structures with the
MSE based ones. So does relay processing matrix. A nested iterative
algorithm jointly optimizing the source precoding and relay
processing is proposed. Simulation results show that the proposed
algorithm provides better performance than the existing methods.
%


\bibliography{mybib}
\bibliographystyle{ieeetr}
%

\end{document}